# Multi-Electrode Dielectric Barrier Discharge Actuators: Geometrical Optimization of High Power Density Array

Anthony Tang[1], Alexander Mamishev[2], Igor Novosselov[1,3,*]

[1]*Department of Mechanical Engineering, University of Washington, Seattle, U.S.A. 98195*

[2]*Department of Electrical and Computer Engineering, University of Washington, Seattle, U.S.A. 98195*

[3]*Institute for Nano-Engineered Systems, University of Washington, Seattle, U.S.A. 98195*

## ABSTRACT

Dielectric barrier discharge (DBD) plasma actuator arrays have been suggested as active flow control devices due to the robust electrohydrodynamic (EHD) force generation in variable atmospheric conditions. DBD plasma augmentation schemes allow for significant performance improvements. However, the transitions to sliding discharge or counter-flow discharge limit their use in high-power arrays. Here, we experimentally demonstrate the performance of a scalable DBD array for two alternating phases of air-exposed electrode configuration. Plasma emissions, direct thrust, velocity profiles, and power consumption measurements of the DBD array reveal that cross-talk between DBD stages can be eliminated to create high-power density actuators. AC augmentation of plasma provides additional gains in thrust; however, the transition to sliding and filamentary discharge reveals geometric limits when increasing the array power density. Introducing a segmented electrode with a resistor delays the onset of adverse sliding and filamentary discharge, allowing it to operate at higher voltage inputs. An optimized four-stage DBD array generated thrust > 250 mN/m with a wall jet thickness > 15 mm, enabling a broader range of flow control applications.

## 1. INTRODUCTION

Non-thermal plasma devices for active flow control have seen great scientific and engineering interest over the past two decades [1-3]. Plasma actuators have the potential to instantaneously influence a fluid system by exerting an electrohydrodynamic (EHD) on the body of fluid while staying silent and compact [4-6]. Categorized by corona discharge or dielectric barrier discharge (DBD), a plasma actuator generates a force by ionizing the fluid around the actuator with an electric field that exceeds the dielectric strength of the working fluid. The free electrons and ions are then accelerated in the electric field, and through collisions with the neutral air molecules, an EHD force is generated. Corona discharge actuators are primarily excited by high voltage direct current (DC), while DBD actuators are excited by high voltage pulsed or alternating current (AC) [7, 8]. Despite lower electromechanical efficiency than corona-driven actuators, DBD actuators are more stable and can provide a more consistent electrohydrodynamic forcing over a range of atmospheric conditions [9-11]. DBD actuators have been used for aerodynamic drag reduction [12, 13], lift augmentation [14, 15], separation control [16, 17], and electric propulsion [18-21]. However, due to their relatively weak EHD forces, current DBD applications are limited to flow control at low-speed conditions [22-24]. To expand plasma actuators' operating range and applications, it is essential to optimize DBD actuator systems [25].

DBD forcing on a fluid volume can be increased by introducing a biased third electrode [26, 27]; both positive and negative biases were previously reported [28, 29]. In this arrangement, the DBD electrode pair ionizes the gas, and the third electrode accelerates positive or negative species, promoting their interaction with neutral molecules. Previous work shows that negative DC-augmentation (DCA) electrode bias leads to modest improvements before the onset of sliding discharge and a counter jet at the DCA electrode, canceling the gains from positive ion acceleration [30]. These opposing wall jets collide, creating a wall-normal thrust [29]. A positive DCA bias monotonically increases thrust by as much as two-fold. While the results of the DCA-DBD are promising, their implementation requires AC and DC power supplies, which is challenging for large-scale DBD arrays. Tang *et al.* [31] recently reported on the AC-augmented (ACA) DBD actuator concept. The in-phase ACA ($\Phi = 0°$) can cause the early transition to reverse (counter-flow) discharge. In comparison, out-of-phase ACA ($\Phi = 180°$) can accelerate both the positive and negative species generated by primary DBD, enhancing momentum transfer to the neutral molecules. The stronger E-field between the two air-exposed DBD electrodes can increase EHD forcing by up to ~ 40%. The authors suggested that the improvements in thrust are primarily due to the additional charge pull-action by the third electrode for both positive and negatively charged species.

---

[*] ivn@uw.edu

Earlier efforts tested DBD arrays constructed of DBD actuators in series, demonstrating increased overall thrust. However, backward discharge (sometimes called a cross-talk phenomenon) is produced on downstream air-exposed electrodes toward previous embedded electrodes when the components are closely spaced. The backward discharge creates a counter-flow, limiting the overall system's efficiency [32-34]. These works tested DBD arrays with different electrode geometries, waveforms, and spacings. Thomas *et al.* [33] noted that the DBD array thrust does not increase linearly with the number of actuator stages. Subsequential DBD array investigations have attempted to reduce the cross-talk by modifying the DBD actuator geometry. Benard *et al.* [35] proposed a DBD array comprised of a three-electrode DBD actuator with two dielectric layers and two embedded electrodes per actuator stage. The authors reduced backward flow due to cross-talk by up to 65%. However, some cross-talk persisted, and the additional embedded electrode increased manufacturing complexity while requiring higher than typical voltages.

More recently, reports have expanded on multiple DBDs in series for continuously accelerating the flow by EHD force; some works demonstrated that wire-to-planar electrodes array with alternating air-exposed HV and grounded electrodes experienced minimal cross-talk. This geometry was first presented by Debien *et al.* [36] with a four-actuator DBD system that created an EHD jet velocity of ~ 10.5 ms$^{-1}$. Sato *et al.* demonstrated linearly scaling DBD array thrust for the first time using custom power electronics [37]. This study used a DBD configuration similar to Debien *et al.* [36] but used planar-to-planar electrodes and the nanosecond pulsed DC voltage [37]. Sato *et al.* introduced a resistor to the downstream air-exposed electrode to prevent arcing to the downstream electrode [37]. This approach continuously accelerates an EHD jet; however, reported results were limited to low-power operations, and only the thrust measurements are presented. The lack of system characterization and impacts of operating conditions on the high-power actuator array performance limit the practical application of the technology.

This study explores a high-power DBD array with AC augmentation (ACA) of DBD plasma discharge at each stage. We present the electrical, mechanical, and plasma characteristics of a DBD array with alternating high-voltage electrodes to create a continuously accelerated boundary layer momentum injection. Effects of geometric variables such as the number of stages and spacing inform the operational limits of DBD arrays. The direct thrust measurements gauge the mechanical performance of the DBD system. At the same time, wall jet velocity profiles allow for the analysis of momentum transfer and mechanisms of the successively accelerated EHD jet. The total electric-to-kinetic energy transfer is evaluated for the various electrical and geometric conditions.

## 2. EXPERIMENTAL SETUP AND DIAGNOSTICS

Traditional metrics to characterize plasma actuators' performance include plasma imaging, mechanical thrust, jet velocity profile, electrical current, and power consumption. Steady-state and time-resolved plasma imaging shed insights into plasma extension, charge density, and the type of discharge. A sensitive and electrically isolated force balance typically measures the mechanical thrust.

### 2.1. DBD Actuator Array

In this study, the two different geometric configurations are tested: (1) an alternating electrode DBD array similar to that in Debien *et al.* [36] and (2) a "resistive" alternating DBD array with an additional electrode connected by a 1 MΩ resistor before each active electrode similar to that in Sato *et al.* [37]. Figure 1 illustrates the two different configurations. A resistive array (with segmented exposed electrodes) limits arcing and filamentary streamers. Since the DBD stages have staggered phases, embedded electrodes experience the same voltage as the following active electrode, preventing reverse discharge. For this manuscript, the alternating electrode DBD array and resistive alternating DBD array will be referred to as the DBD array and the RDBD array. The electromechanical characteristics of the RDBD array were experimentally found to be nearly identical when the varying resistor from 500 kΩ and 10 MΩ. The DBD array is tested with a spacing between each stage L = 20 mm ($DBD_{L=20}$), and the RDBD is tested with spacing between each stage L = 20 mm ($RDBD_{L=20}$) and 10 mm ($RDBD_{L=10}$). The DBD array with L = 10 mm ($DBD_{L=10}$) was not characterized due to strong filamentary streamers. However, the plasma emission of the $DBD_{L=10}$ is discussed. Each configuration is tested with four DBD stages. Since there are several air-exposed and embedded electrodes, an active electrode will be referred to as any air-exposed electrode.

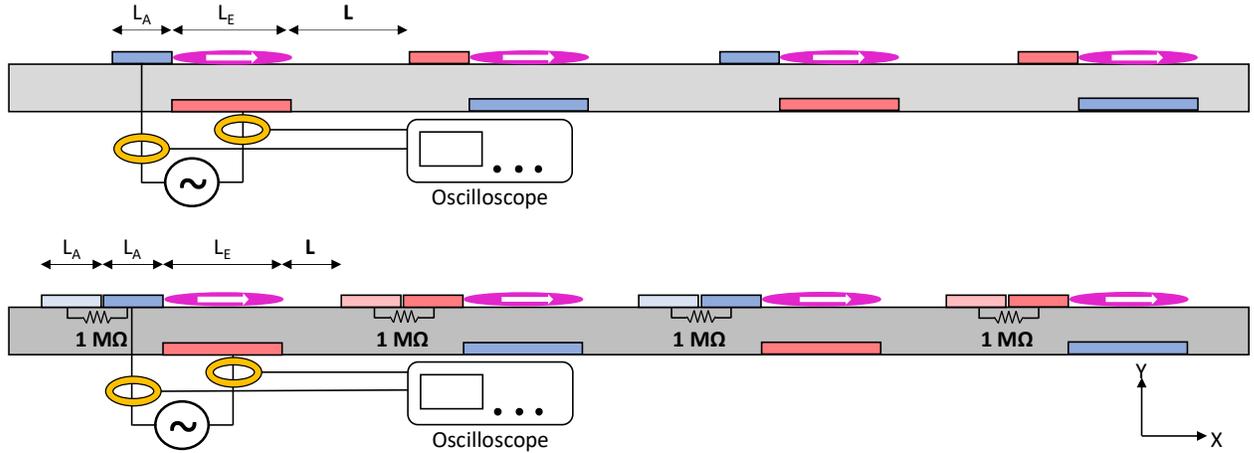

**Figure 1.** The schematic of the alternating electrode DBD array (top) and "resistive" alternating electrode DBD array (bottom) with N = 4 stages. $L_A$ is the length of an air-exposed electrode. $L_E$ -- embedded electrode, and L is the spacing between the DBD stages. $L_A$ and $L_E$ are fixed: $L_A$ = 10 mm and $L_E$ = 20 mm. L varies from 10 mm to 20 mm. The electrodes are powered by a custom power supply with two out-of-phase high-voltage outputs similar to [31, 33]. The electrodes in blue are connected to one of the high-voltage outputs, while the alternating electrodes (red) are connected to the second high-voltage output. In the resistive DBD array configuration, an additional electrode is placed upstream of each air-exposed electrode connected by a 1MΩ resistor.

The dielectric barrier discharge is generated with a high-voltage AC between the active and embedded electrodes. A 3.175 mm quartz dielectric plate separates the active and embedded electrodes. The active electrodes (0.07 mm thick, 10 mm long, and 80 mm wide) are constructed of conductive copper tape and are flush-mounted directly to the top of the dielectric surface. The embedded electrodes (0.07 mm thick, 20 mm long, and 80 mm wide) are also constructed of conductive copper tape and are flush-mounted directly to the backside of the dielectric surface. The embedded electrode is encapsulated with a thick 2.5 mm polyimide/Kapton layer (~7700 VPM, at 25°C) to ensure that backside discharge does not occur. There is no overlap between the active and embedded electrodes. All tests were conducted in quiescent atmospheric pressure air at T ~ 20°C, atmospheric pressure, and 40% - 60% relative humidity.

The active and embedded electrodes are connected to a custom high-voltage power supply similar to Duong et al. [13] and Thomas et al. [33]. Briefly, the custom power supply is comprised of a signal generator (Siglent 1032X), a power amplifier (Crown XLi 3500), and two custom high-voltage transformers (Corona Magnetics). Each transformer produces a maximum voltage ($V_{max}$) of 35 kV and a maximum current ($I_{max}$) of 0.4 A, significantly higher than commercial high voltage amplifiers and necessary for large DBD arrays. The high-voltage output of each transformer is arranged out-of-phase from one another for effectively up to 70 kV peak-to-peak ($V_{pp}$). The two alternating sets of electrodes are each connected to one of the two high-voltage outputs. Mechanical and electrical power measurements show a DBD actuator's performance is identical when powered by a single transformer and electrical ground or two transformers for the same potentials between the electrodes. All tests in this study use a fixed frequency *(f)* of 2 kHz sine wave with voltage across the active and embedded electrode up to $V_{pp}$ = 45 kV. Previous studies have shown that a sine waveform is efficient for DBD actuation with an optimal frequency *f* ~ 1.5 - 4 kHz [38-40]. The feed voltage is monitored with a Tektronix P6015A high-voltage probe.

## 2.2. Thrust and Wall Jet Characterization

*Thrust measurements* are a direct gauge for the mechanical performance of DBD systems, and wall jet measurements allow for fluid analysis of the successively accelerated EHD jet. The direct thrust is measured by vertically holding the actuator system onto a sensitive force balance [33]. In this configuration, wall-parallel plasma-induced flow is directed away from the balance, and the balance measures the reactive downward thrust. The direct forces were measured with an Ohaus SPX622 analytical balance (0 – 620 g range with 0.0098 mN resolution). We placed a Faraday cage around the scale to prevent electromagnetic interference (EMI) or electrostatic forcing on the scale; the stand holding the test item rested on a non-conductive space that transferred the force to the balance. Thin conductive DBS (Drawn Brazed Strand) wire (0.1 mm outer diameter) connected the power supply to the electrodes to ensure no tension in the thrust measurements [20]. All thrust data points were taken after the DBD thrust reached a steady state (typically < 2 seconds). The readings were transferred from the scale to a computer through an electrically isolated USB cable, and each test condition comprises an average of three test intervals of 10 seconds.

*Velocity measurements* are commonly obtained by custom non-conductive pitot tubes, floating hot-wire anemometry, laser Doppler anemometry (LDA), and particle imaging velocimetry (PIV). These measurements characterize momentum transferred from charged species to the surrounding fluid. High temporal resolution velocity measurements have shown that both the positive and negative voltage half-cycles contribute to an EHD force, and their relevant contributions are the topic of active scientific discussions [41, 42]. To measure and characterize the flow field, we employ a custom-made glass pitot tube with a 0.4 mm inner diameter and 0.8 mm outside diameter to measure the time-averaged x-velocity profile identical to previous work [40]. Compared to traditional stainless steel pitot tubes, the glass tube minimizes electrical interaction with the discharge. This method has been extensively used to characterize plasma actuators' performance [36, 39, 43]. The pitot tube is mounted on an optical table and controlled on the x and y-axis by linear stages connected to an Ashcroft CXLdp differential pressure transmitter (0 – 25 Pa with 0.25% accuracy). The pressure transducer outputs a 4 – 20 mA current, linear in its pressure range, placed in series with a 1.5 kΩ resistor. The pressure within the pitot tube equilibrated nearly instantly after changing the flow condition. The voltage across the resistor is recorded for at least 20 seconds with a Hydra Data Logger II. With the time-averaged pressure ($P$), a time-averaged wind velocity ($v$) is calculated using Bernoulli's equation with a calibration correction factor ($C$) that is characteristic of a custom pitot tube expressed as

$$\Delta P = C\rho v^2, \quad (1)$$

where $\rho$ is the fluid density. In our experiments, the typical velocity measurements had a standard deviation of less than 0.02 m s$^{-1}$ over the sampling period. X-velocity measurements are taken at varying x and y positions downstream and upstream on the active electrode edge. At each x-position, the y-velocity profile is obtained from the surface to 20 mm above the plate at increments of 0.25 mm or 0.5 mm (at higher positions). The streamwise measurements were taken by holding a constant y-position and spanning in the x-direction at 0.5 mm intervals to complete the datasets over a regular measurement grid. Due to the pitot tube dimension, we could not capture velocity at y < 0.4 mm. We assume the velocity is linear between the no-slip condition at y = 0 mm and the data at y = 0.4 mm for plotting purposes.

The DBD actuator's momentum and mechanical power can be calculated from the wall jet velocity profiles through control volume analysis. Considering a 2D control volume (with a spanwise unit length), the integration of the square of the x-velocity along a vertical profile of the DBD actuator reflects the horizontal fluid momentum, $M$, defined as:

$$M = \rho \int_{y=0}^{y=h} U^2(y) dy. \qquad (2)$$

where $U(y)$ is the measured x-velocity at varying heights (y) at a constant x-position. With no external fluid flow before the DBD actuator, a momentum integral after each DBD stage reflects the total force produced after each DBD stage. This method to calculate the EHD force, $F_{EHD}$, of a DBD actuator by integrating velocity profiles confirms the directly measured force balance. Dursher and Roy [44] evaluated the two techniques. They found them in agreement as long as the control volume outside the EHD acceleration region was sufficiently large and captured all the entrained flow. Based on the plasma extension measurements and previous work characterizing the EHD forcing region [40, 45], we evaluate the x-momentum of each DBD stage at 15 mm past each active electrode. From the vertical velocity profile, the mechanical power of the system ($W_{mech}$) can also be computed as

$$W_{mech} = \frac{1}{2} \rho L \int_{y=0}^{y=\infty} U^3(y) dy. \qquad (3)$$

### 2.3. Electrical and Optical Measurements

A DBD system can be electrically characterized by its current and power consumption. A current monitor or an in-line capacitor that can determine the Lissajous curve can evaluate the power consumption. In this work, the total power consumption is derived from the current measurement; these were obtained with a Rogowski coil, which is a superposition of capacitive current, discharge current, and noise. The discharge current comprises numerous peaks in the positive-going cycle due to streamer propagation with the addition of glow discharge during the negative-going cycle [46]. Previous reports characterized the relationship between DBD discharge, capacitance, power consumption, and DBD performance [47, 48]. The discharge current is associated with plasma microdischarges, appearing as a series of short current pulses [36]. The current was measured using a 200 MHz bandwidth non-intrusive Pearson 2877 current monitor with a rise time of 2 ns on each HV output of the two transformers. The current monitor can be placed on either or both high-voltage wires, and it has been confirmed to produce the same current and power readings. The current and voltage monitor is connected to a Siglent SDS 2354X oscilloscope with a sampling rate of 1 GS/s to resolve up to 500 MHz based on the Nyquist sampling theorem. With the voltage and current, the time-averaged electrical power is computed as

$$W_{elec} = f_{AC} \int_{t^*=0}^{t^*=1} V(t) * I(t) \, dt, \qquad (4)$$

where $f_{AC}$ is the frequency of the applied voltage, and $V(t)$ and $I(t)$ are the voltage and current at each point in a voltage period. The normalized time ($t^*$) represents a single cycle. We compute the averaged total electrical power from at least ten voltage periods to reduce the noise impact. With electrical and mechanical power, the total efficiency of the DBD array can be computed as

$$\eta_{total} = \frac{W_{mech}}{W_{elec}}. \qquad (5)$$

Previous work suggests that two-electrode DBD actuators typically have efficiencies < 0.1% [39]. Dimensioned force efficiency metric is also used to quantify the DBD performance, e.g., Debien *et al.* [36], Kreigseis *et al.* [47], Thomas *et al.* [33], and Hoskinson *et al.* [49] have presented force-power diagrams for single two-electrode DBD actuators over a range of frequencies and electrode geometries, and have found DBD actuator thrust to scale approximately linearly with power usage. The dimensioned force efficiency is commonly expressed as

$$\eta_{force} = \frac{F_{EHD}}{W_{elec}}. \qquad (6)$$

The DBD plasma emissions are captured to visualize the DBD actuator array. The DBD array plasma emissions are captured using a Nikon D750 DSLR camera with a Nikon AF-S NIKKOR 70-200 mm f/4G ED VR Zoom lens. The camera is operated with an exposure time of 10 ms to capture the discharge emissions of 20 total discharge cycles.

## 3. RESULTS

### 3.1. Plasma Characteristics

This section discusses the plasma emission characteristics of a four-stage DBD and RDBD array with and without alternating electrodes (Figure 2). When the active electrodes are in phase $\Phi = 0°$ (Figure 2.b), the active electrodes create a forward and reverse DBD [31, 33]. The reverse DBD discharge toward the common embedded electrode contributes to adverse cross-talk in the array. For alternating electrode configuration ($\Phi = 180°$) (Figure 2.c), the cross-talk is effectively eliminated due to negligible potential difference between an upstream embedded electrode and a downstream active electrode. Significantly, each active electrode still experiences a potential difference between its downstream embedded electrode to create a unidirectional plasma extension [31, 37]. While the $DBD_{L=10}$ with alternating-phase active electrodes demonstrates a reduction in cross-talk compared to the traditional in-phase DBD array, eventual sliding and filamentary streamer discharge begins to occur at approximately $V_{p-p} > 40$ kV. Typically, sliding discharge is only seen in three-electrode DC-augmented DBD cases [30], and until recently, sliding discharge has not been reported in DBD array studies. The sliding discharge reflects an extension of the charges on the dielectric surface to the opposite-phase active electrode. In this case, the horizontal thrust typically decreases in the sliding discharge as the ions more readily "slide" to the following active electrode without transferring their momentum to neutral molecules [29, 30]. The sliding discharge at the higher voltages suggests that alternating-electrode DBD arrays have a geometric electrode-spacing limit even when cross-talk is minimized. When the spacing is increased to L = 20 mm, the reverse discharge occurs at higher voltages and with less intensity due to the greater spacing (lower E-field). For the alternated electrodes, at L = 20 mm, sliding discharge was not observed for voltages up to 45 kV.

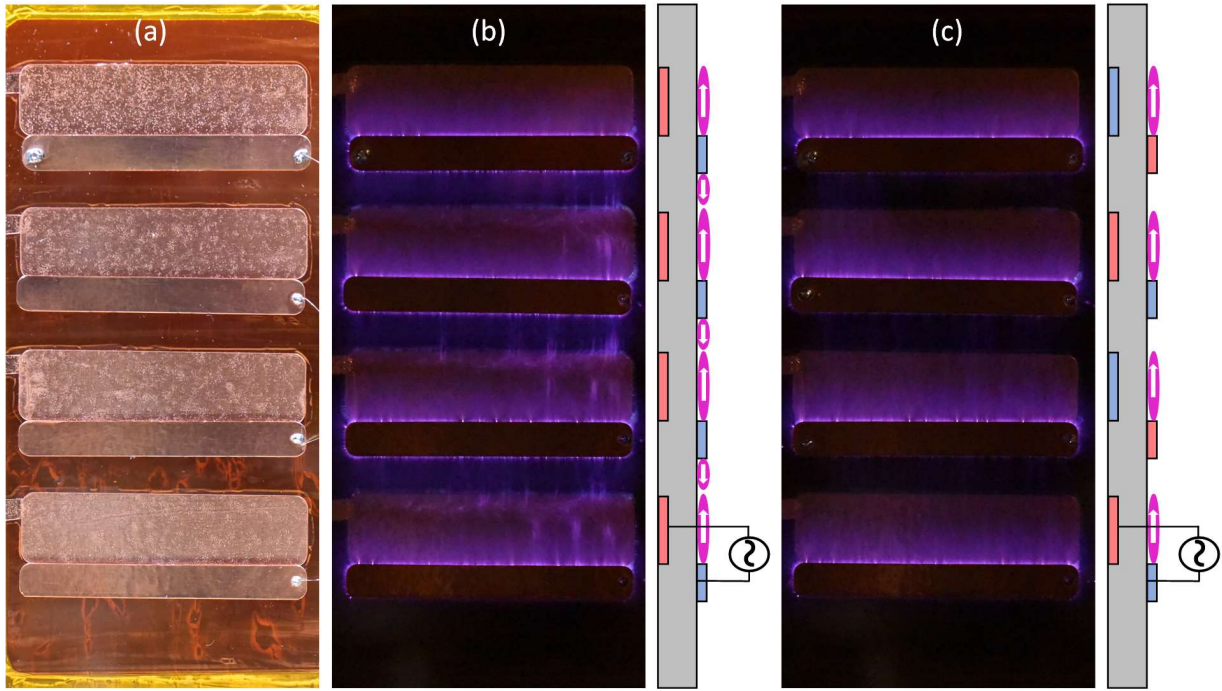

**Figure 2.** Plasma emissions of a four-stage DBD$_{L=10}$ array (a) with in-phase active HV electrodes (b) and alternating-phase active HV electrodes (b). Both DBD arrays are powered at 40 kV / 2 kHz with a 10 ms exposure time to capture the plasma emission of 20 discharge cycles.

Figure 3 shows an RDBD array with in-phase and alternating-phase active electrodes. When an additional electrode with a 1 MΩ resistor is added while maintaining the same L = 10 mm spacing, the RDBD$_{L=10}$ array experiences lower reverse discharge than the DBD$_{L=10}$ array for in-phase configuration (Figure 3.b). For alternating active electrodes (Figure 3.c), sliding and filamentary streamer discharge is delayed until higher voltages (> 45 kV), allowing for the DBD array to operate at higher voltages with greater thrust. The results of the AC RDBD array agree with the observations of the nanopulse DBD array reported by Sato *et al.* [37]. This work is the first to present plasma images of a unidirectional DBD array with a sinusoidal AC waveform. Like the DBD array without a resistor, the RDBD array does not show reverse or sliding discharge L = 20 mm spacing at operating voltages.

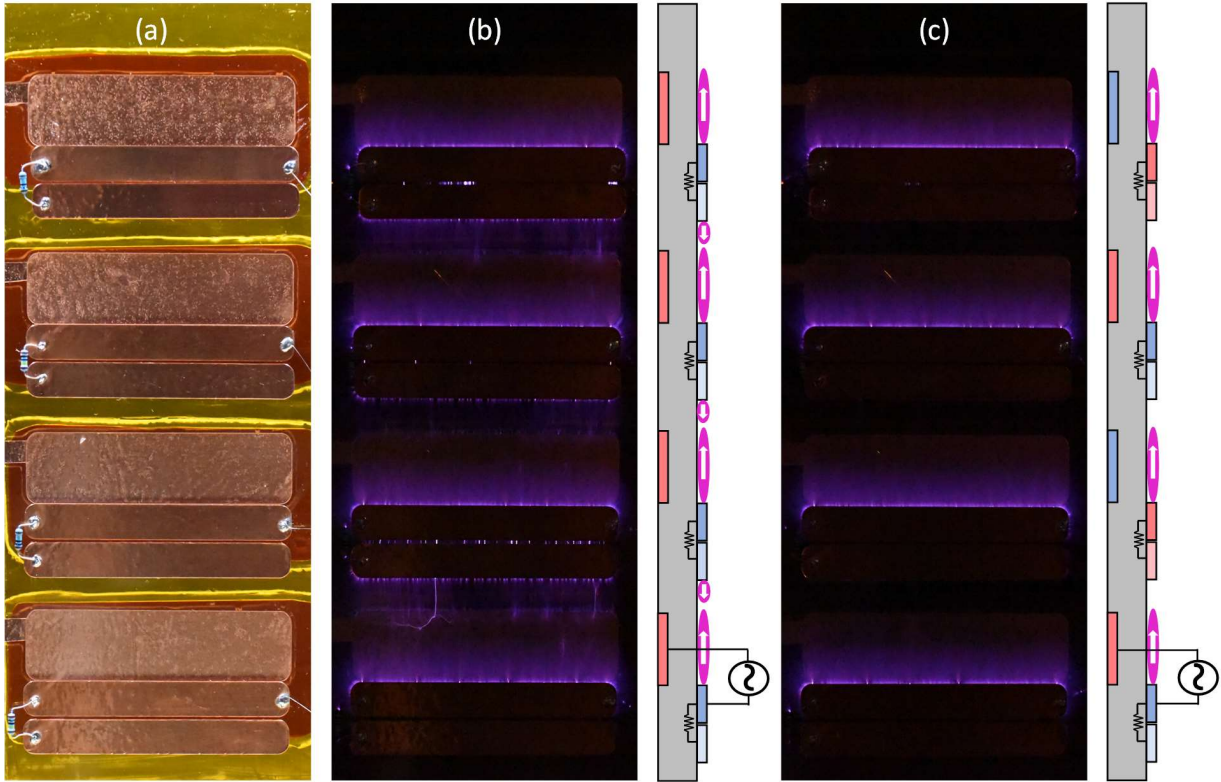

**Figure 3.** Plasma emissions of an $RDBD_{L=10}$ array (a) with air-exposed HV electrodes in series (b) and air-exposed HV electrodes alternating (b). Both DBD arrays are powered at 40 kV / 2 kHz with a 10 ms exposure time to capture the plasma emission of 20 discharge cycles.

### 3.2. DBD Array Thrust Characteristics

The DBD array configurations are found to scale linearly with the number of DBD stages before sliding and filamentary discharge onset. The thrust at varying voltages and number of DBD stages are plotted for the $DBD_{L=10}$, $RDBD_{L=20}$, and $RDBD_{L=10}$ in Figure 4. The linear thrust scaling with up to four DBD stages supports the plasma observations that each DBD stage is similar when the active electrodes are phase-alternating without filamentary streamers or sliding discharge. The $DBD_{L=20}$ array experienced sliding and filamentary streamer discharge at $V_{p-p} \sim 45$ kV; therefore, the $DBD_{L=20}$ array is tested only up to $V_{p-p} = 40$ kV. In comparison, the $RDBD_{L=10}$ array began to exhibit an onset of sliding discharge at $V_{p-p} = 45$ kV. At L = 20 mm, the $RDBD_{L=20}$ array reached a maximum thrust of 251 mN/m with N = 4 stages at 45 kV without the onset of adverse sliding discharge or filamentary streamers. The $DBD_{L=20}$ array reached a maximum thrust of 181 mN/m with N = 4 stages at 40 kV. The $RDBD_{L=20}$ array recorded a thrust of 183 mN/m at the same 40 kV. This suggests that adding the resistor does not significantly improve momentum transfer from each DBD stage and primarily prevents early transition to sliding and filamentary streamer discharge.

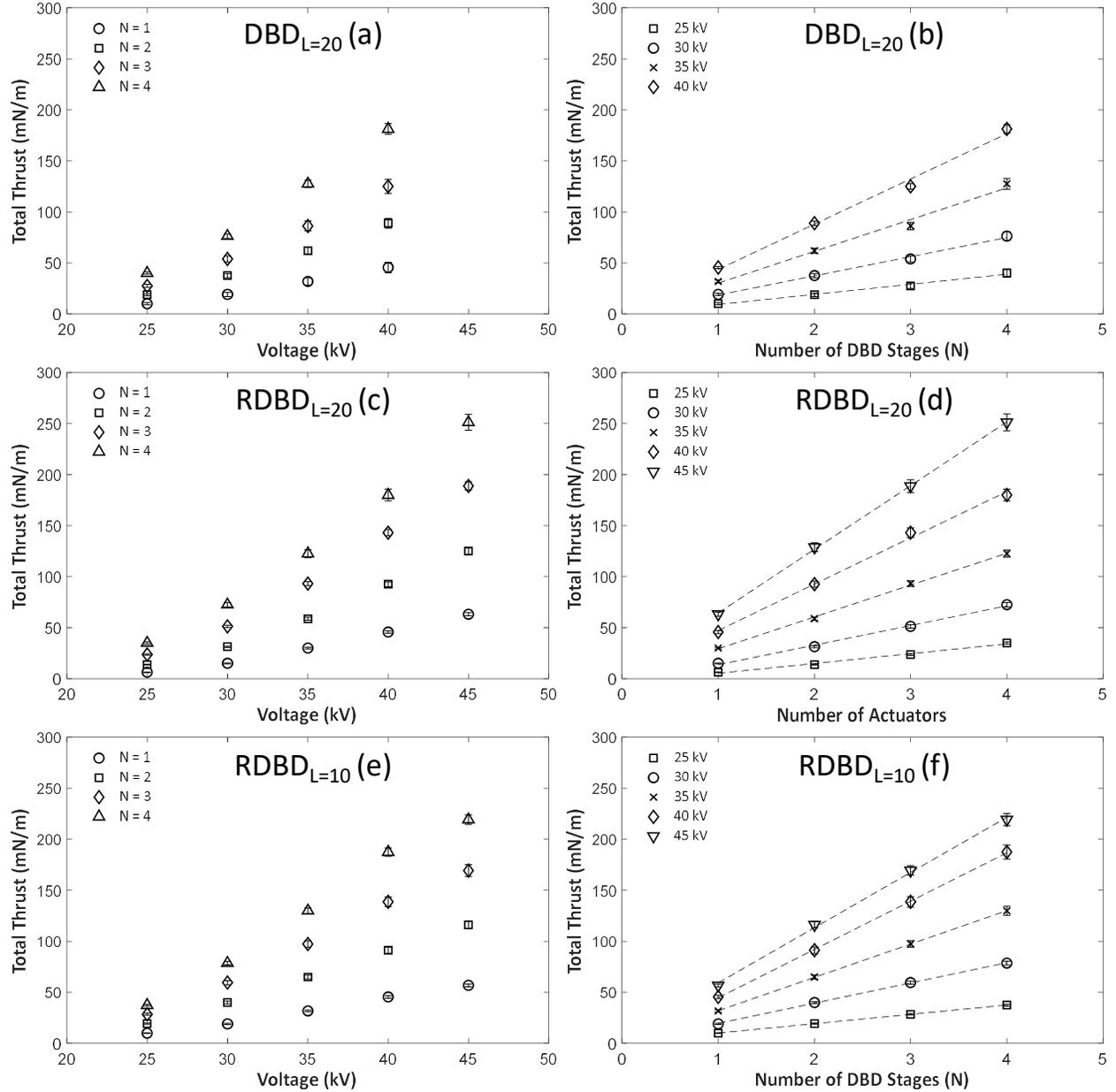

**Figure 4.** $DBD_{L=20}$ array thrust at varying voltages (a) and DBD stages (b), $RDBD_{L=20}$ array thrust at different voltages (c) and DBD stages (d), and $RDBD_{L=10}$ thrust at different voltages (e) and DBD stages (f). Dashlines are provided to illustrate the linear trends. The $DBD_{L=20}$ array shows filamentary streamers at 40 kV, $RDBD_{L=20}$, and the $RDBD_{L=10}$ -- at 45 kV. The standard deviation of the time-averaged thrust across the three sampling periods for each data point is ~ 3%.

For the RDBD array, the maximum thrust at 45 kV decreased from 251 mN/m to 219 mN/m (a 12% decrease) when the spacing was reduced from L = 20 mm to L = 10 mm. While the $RDBD_{L=10}$ at 45 kV did not experience filamentary streamers, the decrease in mechanical performance at this maximum voltage is attributed to the onset of sliding discharges. In this case, the extended ions expelled from one stage reach the sequential air-exposed electrode, limiting momentum transfer to the surrounding air molecules. Before sliding discharges, the $RDBD_{L=10}$ array shows 187 mN/m of total thrust at 40 kV, nearly equal to the thrust of the $RDBD_{L=20}$ array. The similar force before the onset of sliding discharges highlights that the DBD stages produce an EHD force independent of each other at lower voltages with an alternating electrode configuration. However, the proximity of DBD stages to each other is a critical limiting factor due to the transition to sliding discharge.

At the highest tested voltage, $V_{p-p}$ = 45 kV, the $RDBD_{L=20}$ array has a maximum thrust per stage of 62.7 mN/m, while the $RDBD_{L=10}$ has a maximum thrust per stage of 56.5 mN/m. At 40 kV, the simple $DBD_{L=20}$ array has a scaling of 43.8 mN/m, 5% lower than the 46.3 mN/m of the $RDBD_{L=20}$. The improved thrust scaling of the RDBD array with L = 20 mm compared to the same array with L = 10 mm suggests that further spacing between each DBD stage can help with ion acceleration even in the presence of the resistor. Another possible explanation for a slight decrease in performance with the smaller spacing is slight repulsion from the next in-phase HV electrode. For example, ions accelerated from an air-exposed active electrode may experience a wall-normal repulsion from the embedded electrode of the next DBD stage as the two electrodes are in phase with one another. This adverse interaction may be minimized by introducing a gap between each air-exposed electrode and its embedded electrode; further testing or modeling studies are necessary.

### 3.3. Velocity Characteristics

X-velocity profiles of the DBD array across its multiple stages provide insight into the fluid dynamic behavior of the arrays' successive acceleration and linear scaling. Figure 5 shows the x-velocity across the $RDBD_{L=20}$ array at a fixed y-height and fixed x-position on the surface. Figure 6 shows the x-velocity across the $DBD_{L=20}$ array at a fixed y-height and the velocity-derived thrust compared to the directly measured thrust for the two geometries.

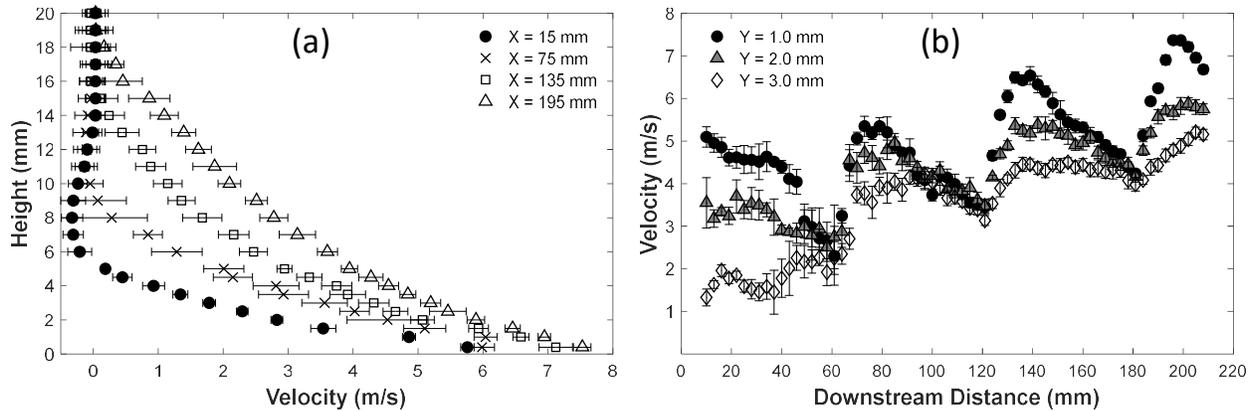

**Figure 5.** X-velocity profile of the $RDBD_{L=20}$ array at 40 kV at varying heights above the dielectric surface at a fixed x-position (a) and varying downstream distances at a fixed y-height (b). The starting x = 0 mm position corresponds to the edge of the first active electrode, and the chosen x-positions for control volume analysis correspond to 15 mm after each active electrode. Two standard deviation error bars are plotted.

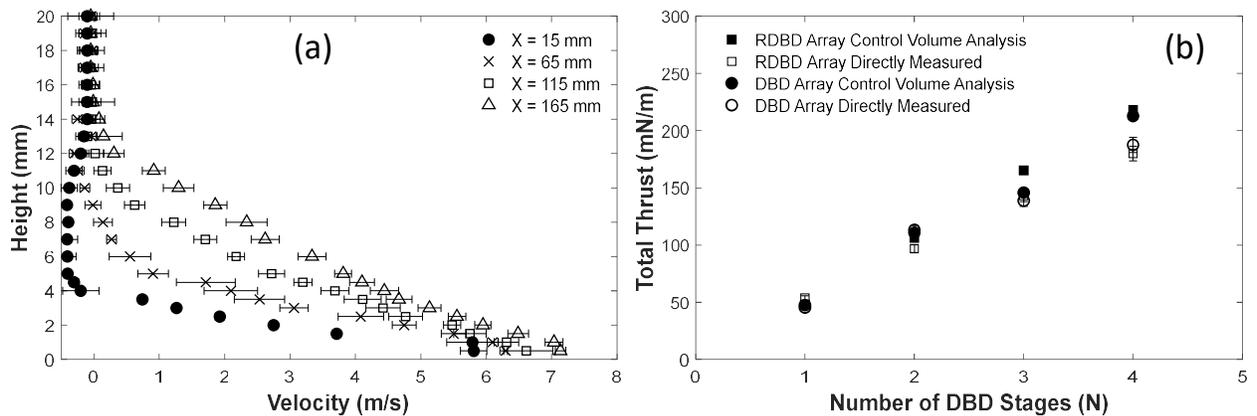

**Figure 6.** X-velocity profile of the $DBD_{L=20}$ array at 40 kV at varying heights above the dielectric surface at a fixed x-position (a). The starting x = 0 mm position corresponds to the edge of the first active electrode, and the chosen x-positions correspond to 15 mm downstream of each active electrode. The velocity-derived thrust using eq.(2) is compared to the directly measured thrust after each DBD stage (b). Two standard deviation error bars are plotted across the > 20 s sampling period.

In both the DBD and RDBD array configuration, the EHD jet after the first stage is greater than 5.5 m/s at the y = 0.4 mm, and the jet increases after each DBD stage up to a maximum velocity after its fourth stage. The $RDBD_{L=20}$ array has a maximum velocity of $V_{max}$= 7.4 m/s, while the simple $DBD_{L=20}$ array recorded a similar $V_{max}$ = 7.1 m/s. Interestingly, both configurations show a slight drop in x-velocity directly above the active electrode of each stage before the subsequent plasma discharge (Figure 5.b). This velocity reduction did not occur in AC-powered three-electrode systems with two exposed electrodes, such as the out-of-phase AC-augmented DBD actuator [31]. This drop in velocity is likely due to the additional mixing of the wall jet and the quiescent air during the momentum injection by the downstream stage. This hypothesis is also supported by the thickening of the wall jet profile at each consecutive stage (Figure 5.a and Figure 6.a). Between the two configurations, the simpler $DBD_{L=20}$ array has a thinner wall jet after the fourth stage due to less mixing over the shorter area.

The velocity profiles are integrated at 15 mm past each active electrode stage to derive the total momentum after each stage (Figure 6.b). The velocity-derived thrust through control volume analysis around each stage is compared to the directly measured thrust. The two methods agree well and within 10%, similar to Dursher and Roy [44]. Additionally, wind tunnel studies on DBD actuators in external co-flow support the linear thrust scaling in a DBD array. Tang *et al.* [6] and Pereira *et al.* [50] demonstrated that a DBD actuator thrust and power consumption is independent of external flow. The scaling of the DBD array is similar to a DBD in external co-flow because each subsequential DBD stage injects momentum into a pre-existing fluid jet. Unlike a typical DBD actuator with a typical wall jet thickness <5 mm, the DBD array is found to produce a wall jet thickness of ~ 20 mm. With a velocity-derived momentum of 202 mN/m at 40 kV at 2 kHz, the wall jet of the multi-stage DBD array in this study's resistive and simple configuration is significantly thicker and with more momentum than currently published literature. While the authors are not aware of any literature demonstrating a wall jet with > 100 mN/m, the $V_{max}$ is lower compared to similar DBD array works, including Debien *et al.* [36], which used a wire-to-plate configuration for a $V_{max}$ = 10.5 m/s at y = 0.6 mm after 4-stage DBD array. The difference may be attributed to the geometric differences, such as a gap between the active wire and embedded electrodes or relatively long active electrodes used in this work. Additional time-resolved flow field measurements and modeling efforts can shed insights into the complex plasma flow interaction in the array.

### 3.4. Electrical Characteristics

The electrical power consumption of the DBD arrays is discussed in this section. Figure 7 displays the total power consumption results derived from the current monitor and eq.(4) with a simple and resistive DBD array.

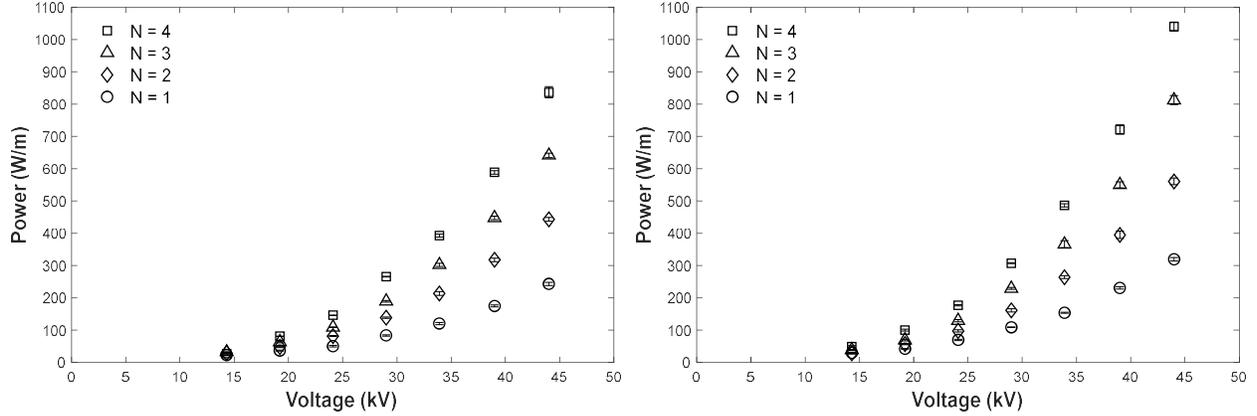

**Figure 7.** DBD Array power usage scaling with a simple $DBD_{L=20}$ array (left) and $RDBD_{L=20}$ array (right) with a fixed L = 20 mm gap. Two standard deviation error bars across a minimum of 10 voltage cycles are plotted. The simple DBD array was tested to 45 kV, however filamentary streamers began to form at the highest voltage.

Like the mechanical thrust performance, the electrical power for the $DBD_{L=20}$ array and the $RDBD_{L=20}$ array scale linearly with the number of stages. Before the onset sliding discharge, the $RDBD_{L=10}$ also demonstrated linear power scaling with the number of DBD stages. For a two-electrode DBD, the power usage approaches 220 W/m at the maximum voltage of 45 kV. It matches the previously reported power usage for single DBD actuators with a similar dielectric thickness and voltage [39]. The maximum power usage of the four-stage $DBD_{L=20}$ array and $RDBD_{L=20}$ array is approximately 840 W/m and 1080 W/m, respectively, at 45 kV. With the same spacing between each DBD stage, the increase in power for the resistive DBD array is attributed to the additional high-voltage electrode with a resistor that also increases the effective area and capacitance within the dielectric. Using Fast-Fourier Transform (FFT) analysis to identify the capacitive current similar to previous works [30, 40], the electrical measurements show that the RDBD array at 40 kV uses ~74% of its total power usage as capacitive power. In contrast, the simple DBD array uses ~64% of its total power as capacitive power.

### 3.5. Force Efficiency and Total Efficiency

With thrust, mechanical power, and electrical power measurements across varying voltages and the number of DBD stages, the force efficiency and total efficiency of the DBD and RDBD array can be calculated. The electrical power measurements of the simple and resistive DBD array are found to scale linearly with the total thrust across the range of DBD stages and at varying voltages, as shown in Figure 8.

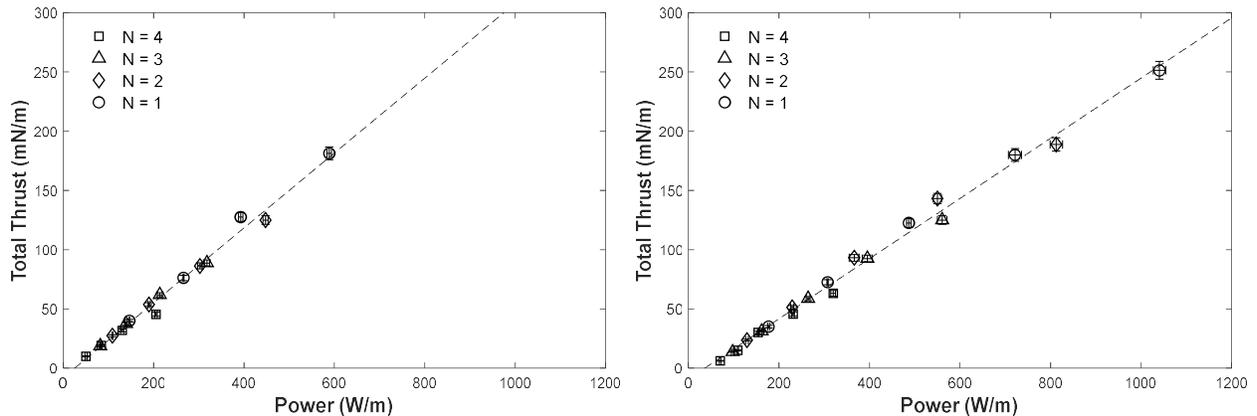

**Figure 8.** Total thrust scaling with power usage of $DBD_{L=20}$ array and $RDBD_{L=20}$. Two standard deviation error bars in power (horizontal) and thrust (vertical) are plotted.

The slope of the thrust-to-power graph represents the force efficiency, $\eta_{force}$, defined in Eq. (6). The simple $DBD_{L=20}$ array is found to have a force efficiency of $\eta_{force}$ = 0.316 mN/W, while the $RDBD_{L=20}$ demonstrated an efficiency of $\eta_{force}$ = 0.254 mN/W. This is the first report of force efficiency for an AC-powered DBD array. The efficiencies agree with previously published AC two-electrode DBD results of $\eta_{force}$ = 0.25 mN/W in Kreigseis et al. [47]. Sato et al. [37] is the only other published work reporting force efficiency ($\eta_{force}$ = 0.06 mN/W) for DBD arrays. Their low efficiency is likely due to the low utilization of the DBD cycle using a pulsed-DC waveform. Greater force efficiency is generally achieved with thicker dielectrics to decrease the system's capacitance, as noted in Jolibois et al. [51]. A summary and comparison of recent force and total efficiencies is presented in Table 1.

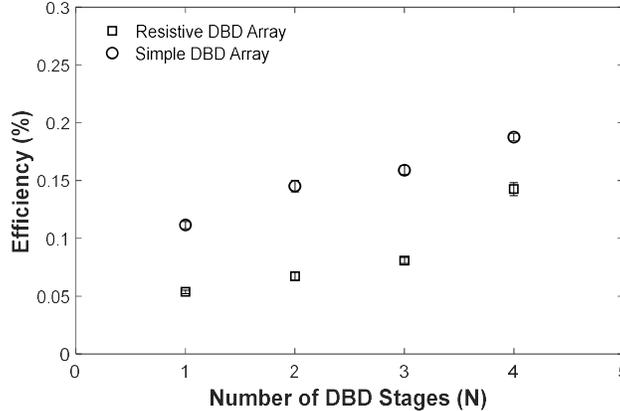

**Figure 9.** Total Efficiency of $DBD_{L=20}$ and $RDBD_{L=20}$ at 40 kV / 2 kHz.

Figure 9 shows the total efficiency of the two DBD arrays with varying DBD stages at 40 kV. Between the simple and resistive arrays, the RDBD array has a lower force efficiency and total efficiency across all numbers of DBD stages at 40 kV. The total efficiency of ~0.1% for a single-stage two-electrode DBD agrees with previous experiments [52, 53]. Interestingly, both DBD array configurations increase in total efficiency with the number of DBD actuator stages. This increase in efficiency with the number of DBD stages is likely due to the AC augmentation mechanism in the alternating (out-of-phase) exposed electrode configuration. For example, our recent work [31] demonstrates improvement in the performance of a planar DBD actuator by utilizing an AC-augmented electrical field in a three-electrode geometry. Time-resolved electrical and optical measurements, velocity profiles, and direct thrust measurements were used to characterize the EHD augmentation. The third electrode, the "pull" action, increased the EHD forcing by up to ~ 40%. The continuous plasma acceleration regions also reduce the viscous losses to the wall.

The RDBD array's lower efficiency than the simple DBD array suggests that the increased power usage due to the added electrode and resistor does not proportionally increase thrust for the same voltages. For example, the $RDBD_{L=20}$ generated approximately the same thrust as the simple $DBD_{L=20}$ at the same voltages before sliding or filamentary discharge. Therefore, since the RDBD array can be operated at a higher voltage, the primary advantage of the RDBD array is its greater maximum operating voltages and thrust. A disadvantage of the RDBD array is the added manufacturing complexity, space requirement, and efficiency. Given a need for efficiency or maximum thrust, each configuration can be uniquely considered for applications requiring a more robust DBD array. Future research efforts should further explore how resistors and other electrical components delay the formation of filamentary streamers and sliding regions, enhancing ion-momentum transfer.

**Table 1.** Summary of force and total efficiency. * Shows the range of efficiencies between 1 and 4 DBD stages.

| | Configuration | Dielectric (thickness) | Waveform | P-P Voltage (kV) | Frequency (kHz) | $\eta_{force}$ | $\eta_{total}$ |
|---|---|---|---|---|---|---|---|
| Presented work | Planar – planar (DBD Array) (RDBD Array) | Quartz (3.175 mm) | Sine | 25 – 45 | 2.0 | 0.25 0.32 | (0.05 - 0.14)* (0.11 - 0.19)* |
| Sato et al. [37] | Planer – planer (RDBD Array) | Polyimide (0.32 mm) | Nanosecond Pulsed DC | 8 | 0.5 – 2.0 | 0.05 – 0.07 | N/A |
| Kriegseis et al. [47] | Planar – planar (Single DBD) | Polyimide (0.4 mm) | Sine | 5 – 15 | 8.0 – 13.0 | 0.25 | N/A |
| Hoskinson and Hershkowitz [54] | Wire – planar (Single DBD) | Polyester (0.25 mm) | Ramp | 12 – 16 | 1.0 | 0.04 – 0.7 | N/A |
| Benard and Moreau [38] | Planar – planar (Single DBD) | PMMA (3 mm) | Sine Square Ramp | 40 | 1.5 | 0.64 0.37 0.22 | N/A N/A N/A |
| Enloe et al. [55] | Planar – planar (Single DBD) | N/A | Sine | 6 – 18 | 5.0 | 0.12 | N/A |
| Debien et al. [36] | Planar – planar (Single DBD) | PMMA (3 mm) | Sine | 24 – 44 | 1.5 | 0.65 | 0.06 – 0.18 |
| Thomas et al. [33] | Planar – planar (Single DBD | Quartz (6.35 mm) | Sine | 20 – 70 | 1.0 – 8.0 | 0.09 – 0.25 | N/A |
| Ferry and Rovey [56] | Planar – planar (Single DBD) | Polyimide (0.2 mm) | Sine | 6 – 9 | 1.0 – 18.0 | 0.02 – 0.1 | N/A |
| Jolibois and Moreau [51] | Planer – planer (Single DBD) | PMMA (3 mm) | Sine, Square Ramp, Trapezoid | 16 – 52 | 1.0 | N/A | 0.01 – 0.09 |

## 4. DISCUSSIONS

Compared to previous works, these results demonstrate linear thrust scaling for high-power density DBD arrays with voltages. Sato et al. [37] demonstrated the linear thrust scaling up to ~80 mN/m with a four-stage DBD array, significantly lower than the shown maximum 251 mN/m thrust with the four-stage RDBD presented here. Thomas et al. [33] varied electrode widths and electrical waveforms and found that a three-stage DBD array (N = 3) experienced, at best, ~2.4x the force of a single-stage DBD actuator, showing diminishing array efficiency.

The compounded effect of the DBD stages on momentum injection and overall trust is analogous to co-flow momentum injections into the existing flow profile. Our recent work shows that for the laminar boundary layer, the momentum injection by the DBD actuator in co-flow is nearly identical to the momentum injection into the quiescent environment [6]. For the DBD array (in laminar flow), each subsequent momentum injection is independent of the previous ones, which allows for building the overall wall jet momentum over the array's length. This work does not address the effect of turbulent mixing, typically present in the free stream flow, on array momentum injection.

The high efficiency of the array is linked to the AC augmentation of the primary DBD discharge in the alternating (out-of-phase) exposed electrode configuration. The mechanism described by Tang et al. [31] shows the benefits of the third electrode, similar to the alternating electrode configuration in this work. The third electrode, the "pull" action, increased the EHD forcing by up to ~ 40%. The continuous plasma acceleration regions also reduce the viscous losses, resulting in higher momentum injection.

## 5. CONCLUSION

In conclusion, this is the first report demonstrating a high-powered scalable DBD array actuator with minimal interference between each DBD stage and thrust greater than 250 mN/m (N = 4). The DBD arrays are tested in two geometries with a conventional AC waveform with up to four alternating-phase DBD stages: (i) a simple DBD array with each DBD stage consisting of the traditional air-exposed electrode and an embedded electrode and (ii) a "resistive" DBD array with each DBD stage consisting of two air-exposed electrodes connected by a 1 MΩ resistor and an embedded electrode. Across multiple voltages and DBD stages, the results show a linear relationship between power consumption and thrust for both DBD array configurations before the onset of adverse sliding or filamentary streamer discharge. The added resistor in the RDBD array was found to delay the onset of sliding and filamentary discharge, allowing for higher voltage operation and greater thrust. However, the simple DBD array was more efficient because the added electrode and resistor increased electrical power consumption without improving momentum transfer from the plasma discharge. With an increase in total efficiency with each additional DBD stage, the presented results allow for more robust applications of DBD actuation with significantly greater forces and efficiencies than previously demonstrated. The limitation of this work includes average velocity measurements that do not allow for the study of the temporal effects of the EHD flow interaction. Future work should consider experiments in the turbulence boundary layer and realistic geometries relevant to active flow control applications.

## NOMENCLATURE

| | |
|---|---|
| $C$ | Pitot tube correction factor |
| $E$ | Electric field |
| $f_{AC}$ | Frequency of the applied voltage |
| $\vec{f}_{EHD}$ | Electrohydrodynamic force term |
| $i(t)$ | Current |
| $I_{dis}$ | Discharge current |
| $L$ | Spanwise Length |
| $M$ | The momentum of the induced jet |
| $P$ | Pressure reading from the pitot tube |
| $W$ | Discharge energy consumption |
| $W_{mech}$ | Mechanical power |
| $W_{elec}$ | Electrical power |
| $U(y)$ | Velocity at y height |
| $U_{max}$ | Maximum velocity of the wall jet |
| $U_\infty$ | External flow velocity |
| $V_{AC}$ | AC Voltage in the DBD actuator |
| $v$ | Time-averaged velocity |
| $t^*$ | Normalized time value |
| $\rho$ | Density |
| $Q$ | Mass flow rate |

# REFERENCES

[1] I. Adamovich *et al.*, "The 2017 Plasma Roadmap: Low temperature plasma science and technology," *Journal of Physics D: Applied Physics,* vol. 50, no. 32, p. 323001, 2017.

[2] E. Moreau, "Airflow control by non-thermal plasma actuators," *Journal of Physics D: Applied Physics,* vol. 40, no. 3, pp. 605-636, 2007, doi: 10.1088/0022-3727/40/3/S01.

[3] A. Tang, A. Ong, N. Beck, J. S. Meschke, and I. V. Novosselov, "Surface Virus Inactivation by Non-Thermal Plasma Flow Reactor," 2023.

[4] T. C. Corke, M. L. Post, and D. M. Orlov, "SDBD plasma enhanced aerodynamics: concepts, optimization and applications," *Progress in Aerospace Sciences,* vol. 43, no. 7, pp. 193-217, 2007, doi: 10.1016/j.paerosci.2007.06.001.

[5] T. C. Corke, C. L. Enloe, and S. P. Wilkinson, "Dielectric Barrier Discharge Plasma Actuators for Flow Control *," *Annu. Rev. Fluid Mech.,* vol. 42, no. 1, pp. 505-529, 2010, doi: 10.1146/annurev-fluid-121108-145550.

[6] A. Tang, N. Li, B. Price, A. Mamishev, A. Aliseda, and I. Novosselov, "Dielectric barrier discharge actuators: Momentum injection into co-flow and counter-flow freestream," *Journal of Electrostatics,* vol. 129, p. 103918, 2024/06/01/ 2024, doi: https://doi.org/10.1016/j.elstat.2024.103918.

[7] G. Touchard, "Plasma actuators for aeronautics applications-State of art review," *International Journal of Plasma Environmental Science and Technology,* vol. 2, no. 1, pp. 1-25, 2008.

[8] Z. Su, H. Zong, H. Liang, J. Li, and Y. Wu, "Optimization in frequency characteristics of an oscillating dielectric barrier discharge plasma actuator," *Sensors and Actuators A: Physical,* vol. 351, p. 114195, 2023.

[9] T. Orrière, É. Moreau, and D. Z. Pai, "Electric wind generation by nanosecond repetitively pulsed microplasmas," *Journal of physics. D, Applied physics,* vol. 52, no. 46, p. 464002, 2019, doi: 10.1088/1361-6463/ab372f.

[10] P. Warlitz, M. T. Hehner, S. Pasch, J. Serpieri, T. Blank, and J. Kriegseis, "On durable materials for dielectric-barrier discharge plasma actuators," *Sensors and Actuators A: Physical,* vol. 366, p. 114985, 2024.

[11] S. Sato *et al.*, "Fabrication and performance evaluation of full-inkjet-printed dielectric-barrier-discharge plasma actuators," *Sensors and Actuators A: Physical,* vol. 344, p. 113751, 2022.

[12] M. T. Hehner, D. Gatti, and J. Kriegseis, "Stokes-layer formation under absence of moving parts—A novel oscillatory plasma actuator design for turbulent drag reduction," *Physics of Fluids,* vol. 31, no. 5, p. 051701, 2019.

[13] A. Duong, T. Corke, F. Thomas, and K. Yates, "Turbulent boundary layer drag reduction using pulsed-dc plasma actuation," *Journal of Fluid Mechanics,* 2019.

[14] M. Han, J. Li, Z. Niu, H. Liang, G. Zhao, and W. Hua, "Aerodynamic performance enhancement of a flying wing using nanosecond pulsed DBD plasma actuator," *Chinese Journal of Aeronautics,* vol. 28, no. 2, pp. 377-384, 2015, doi: 10.1016/j.cja.2015.02.006.

[15] N. Benard, J. Jolibois, and E. Moreau, "Lift and drag performances of an axisymmetric airfoil controlled by plasma actuator," *Journal of Electrostatics,* vol. 67, no. 2 3, p. 133, 2009, doi: 10.1016/j.elstat.2009.01.008.

[16] S. G. Pouryoussefi, M. Mirzaei, F. Alinejad, and S. M. Pouryoussefi, "Experimental investigation of separation bubble control on an iced airfoil using plasma actuator," *Applied Thermal Engineering,* vol. 100, pp. 1334-1341, 2016.

[17] L. Francioso, C. De Pascali, E. Pescini, M. De Giorgi, and P. Siciliano, "Modeling, fabrication and plasma actuator coupling of flexible pressure sensors for flow separation detection and control in aeronautical applications," *Journal of Physics D: Applied Physics,* vol. 49, no. 23, p. 235201, 2016.

[18] H. Xu *et al.*, "Flight of an aeroplane with solid-state propulsion," *Nature,* vol. 563, no. 7732, pp. 532-535, 2018.

[19] D. S. Drew and K. S. Pister, "First takeoff of a flying microrobot with no moving parts," in *Manipulation, Automation and Robotics at Small Scales (MARSS), 2017 International Conference on*, 2017: IEEE, pp. 1-5.

[20] H. K. Hari Prasad, R. S. Vaddi, Y. M. Chukewad, E. Dedic, I. Novosselov, and S. B. Fuller, "A laser-microfabricated electrohydrodynamic thruster for centimeter-scale aerial robots," *PloS one,* vol. 15, no. 4, p. e0231362, 2020.

[21] R. S. Vaddi, Y. Guan, A. Mamishev, and I. Novosselov, "Analytical model for electrohydrodynamic thrust," *Proceedings A: Mathematical, Physical and Engineering Sciences,* vol. 476, no. 2241, 2020.

[22] D. Keisar, D. Hasin, and D. Greenblatt, "Plasma actuator application on a full-scale aircraft tail," *AIAA Journal,* vol. 57, no. 2, pp. 616-627, 2019.

[23] H. Akbıyık, H. Yavuz, and Y. E. Akansu, "Comparison of the linear and spanwise-segmented DBD plasma actuators on flow control around a NACA0015 airfoil," *IEEE Transactions on Plasma Science,* vol. 45, no. 11, pp. 2913-2921, 2017.

[24] P. G. Eijo, R. Sosa, J.-L. Aider, and J. M. Cabaleiro, "Control of air micro-jets by the use of dielectric barrier discharges," *Sensors and Actuators A: Physical,* vol. 305, p. 111937, 2020.

[25] R. S. Vaddi, "Electrohydrodynamic Actuators for Propulsion and Flow Control," 2021.

[26] B.-R. Zheng, M. Xue, and C. Ge, "Dynamic evolution of vortex structures induced by tri-electrode plasma actuator," *Chinese physics B,* vol. 29, no. 2, p. 24704, 2020, doi: 10.1088/1674-1056/ab671f.

[27] K. Chen, X. Geng, Z. Shi, K. Cheng, and H. Cui, "Experimental investigation of influence of sliding discharge DBD plasma on low-speed boundary layer," *AIP advances,* vol. 10, no. 3, pp. 35108-035108-9, 2020, doi: 10.1063/1.5134848.

[28] E. Moreau, R. Sosa, and G. Artana, "Electric wind produced by surface plasma actuators: a new dielectric barrier discharge based on a three-electrode geometry," *Journal of Physics D: Applied Physics,* vol. 41, no. 11, p. 115204, 2008.

[29] A. Debien, N. Benard, and E. Moreau, "Electric wind produced by sliding discharges," *Proceeding of 2nd ISNPEDADM new electrical technologies for environment, Nouméa,* 2011.

[30] A. Tang, A. Aliseda, A. Mamishev, and I. Novosselov, "DC-Augmented Dielectric Barrier Discharge (DCA-DBD)," *arXiv preprint arXiv:2403.18064,* 2024.

[31] A. Tang, A. Mamishev, and I. Novosselov, "AC-Augmented Dielectric Barrier Discharge," *arXiv preprint arXiv:2411.17677,* 2024.

[32] M. Forte, J. Jolibois, J. Pons, E. Moreau, G. Touchard, and M. Cazalens, "Optimization of a dielectric barrier discharge actuator by stationary and non-stationary measurements

of the induced flow velocity: application to airflow control," *Experiments in Fluids,* vol. 43, no. 6, pp. 917-928, 2007, doi: 10.1007/s00348-007-0362-7.

[33] F. O. Thomas, T. C. Corke, M. Iqbal, A. Kozlov, and D. Schatzman, "Optimization of Dielectric Barrier Discharge Plasma Actuators for Active Aerodynamic Flow Control," *AIAA Journal,* vol. 47, no. 9, pp. 2169-2178, 2009, doi: 10.2514/1.41588.

[34] A. Berendt, J. Podliński, and J. Mizeraczyk, "Elongated DBD with floating interelectrodes for actuators," *The European Physical Journal-Applied Physics,* vol. 55, no. 1, p. 13804, 2011.

[35] N. Benard, A. Mizuno, and E. Moreau, "A large-scale multiple dielectric barrier discharge actuator based on an innovative three-electrode design," *Journal of physics. D, Applied physics,* vol. 42, no. 23, p. 235204, 2009, doi: 10.1088/0022-3727/42/23/235204.

[36] A. Debien, N. Benard, and E. Moreau, "Streamer inhibition for improving force and electric wind produced by DBD actuators," *Journal of Physics D: Applied Physics,* vol. 45, no. 21, p. 215201, 2012.

[37] S. Sato, H. Furukawa, A. Komuro, M. Takahashi, and N. Ohnishi, "Successively accelerated ionic wind with integrated dielectric-barrier-discharge plasma actuator for low-voltage operation," *Scientific reports,* vol. 9, no. 1, pp. 1-11, 2019.

[38] N. Benard and E. Moreau, "Role of the electric waveform supplying a dielectric barrier discharge plasma actuator," *Applied Physics Letters,* vol. 100, no. 19, p. 193503, 2012.

[39] N. Benard and E. Moreau, "Electrical and mechanical characteristics of surface AC dielectric barrier discharge plasma actuators applied to airflow control," *Experiments in Fluids,* vol. 55, no. 11, p. 1846, 2014.

[40] A. Tang, R. S. Vaddi, A. Mamishev, and I. V. Novosselov, "Empirical relations for discharge current and momentum injection in dielectric barrier discharge plasma actuators," *Journal of Physics D: Applied Physics,* vol. 54, no. 24, p. 245204, 2021/03/31 2021, doi: 10.1088/1361-6463/abec0b.

[41] I. Biganzoli, R. Barni, and C. Riccardi, "Temporal evolution of a surface dielectric barrier discharge for different groups of plasma microdischarges," *Journal of physics. D, Applied physics,* vol. 46, no. 2, p. 025201, 2013, doi: 10.1088/0022-3727/46/2/025201.

[42] N. Benard, P. Noté, M. Caron, and E. Moreau, "Highly time-resolved investigation of the electric wind caused by surface DBD at various ac frequencies," *Journal of electrostatics,* vol. 88, pp. 41-48, 2017, doi: 10.1016/j.elstat.2017.01.018.

[43] J. Pons, E. Moreau, and G. Touchard, "Electrical and aerodynamic characteristics of atmospheric pressure barrier discharges in ambient air," 2004.

[44] R. Durscher and S. Roy, "Evaluation of thrust measurement techniques for dielectric barrier discharge actuators," *Experiments in fluids,* vol. 53, pp. 1165-1176, 2012.

[45] A. Debien, N. Benard, L. David, and E. Moreau, "Unsteady aspect of the electrohydrodynamic force produced by surface dielectric barrier discharge actuators," *Applied Physics Letters,* vol. 100, no. 1, 2012.

[46] C. Louste, G. Artana, E. Moreau, and G. Touchard, "Sliding discharge in air at atmospheric pressure: electrical properties," *Journal of Electrostatics,* vol. 63, no. 6-10, pp. 615-620, 2005.

[47] J. Kriegseis, S. Grundmann, and C. Tropea, "Power consumption, discharge capacitance and light emission as measures for thrust production of dielectric barrier discharge

plasma actuators," *Journal of Applied Physics,* vol. 110, no. 1, 2011, doi: 10.1063/1.3603030.

[48] M. Kuhnhenn, B. Simon, I. Maden, and J. Kriegseis, "Interrelation of phase-averaged volume force and capacitance of dielectric barrier discharge plasma actuators," *Journal of Fluid Mechanics,* vol. 809, 2016.

[49] A. R. Hoskinson and N. Hershkowitz, "Differences between dielectric barrier discharge plasma actuators with cylindrical and rectangular exposed electrodes," *Journal of Physics D: Applied Physics,* vol. 43, no. 6, p. 065205, 2010, doi: 10.1088/0022-3727/43/6/065205.

[50] R. Pereira, D. Ragni, and M. Kotsonis, "Effect of external flow velocity on momentum transfer of dielectric barrier discharge plasma actuators," *Journal of applied physics,* vol. 116, no. 10, p. 103301, 2014, doi: 10.1063/1.4894518.

[51] J. Jolibois and E. Moreau, "Enhancement of the electromechanical performances of a single dielectric barrier discharge actuator," *IEEE Transactions on Dielectrics and Electrical Insulation,* vol. 16, no. 3, pp. 758-767, 2009.

[52] M. Kotsonis and S. Ghaemi, "Performance improvement of plasma actuators using asymmetric high voltage waveforms," *Journal of Physics D: Applied Physics,* vol. 45, no. 4, p. 045204, 2012.

[53] R. Giepman and M. Kotsonis, "On the mechanical efficiency of dielectric barrier discharge plasma actuators," *Applied Physics Letters,* vol. 98, no. 22, 2011.

[54] A. Hoskinson, N. Hershkowitz, and D. Ashpis, "Comparisons of force measurement methods for DBD plasma actuators in quiescent air," in *47th AIAA Aerospace Sciences Meeting including The New Horizons Forum and Aerospace Exposition*, 2009, p. 485.

[55] C. L. Enloe, T. E. McLaughlin, G. I. Font, and J. W. Baughn, "Parameterization of Temporal Structure in the Single-Dielectric-Barrier Aerodynamic Plasma Actuator," *AIAA Journal,* vol. 44, no. 6, pp. 1127-1136, 2006, doi: 10.2514/1.16297.

[56] J. Ferry and J. Rovey, "Thrust measurement of dielectric barrier discharge plasma actuators and power requirements for aerodynamic control," in *5th Flow Control Conference*, 2010, p. 4982.